\documentstyle[12pt]{article}
\newcommand{\s}{\mbox{$\Sigma_{g}$}}
\newcommand{\ZJ}{\mbox{$Z_{\Sigma_{g}}(J)$}}
\newcommand{\J}{\mbox{$\bf J$}}

\newcommand{\y}{\mbox{$\int_{\Sigma_{g}}d\mu$}}
\newcommand{\yy}{\mbox{$\int_{\Sigma_{g}}$}}
\newcommand{\p}{\mbox{$\phi$}}
\newcommand{\PP}{\mbox{$\bf \phi$}}
\newcommand{\T}{\mbox{\rm Tr}}
\newcommand{\e}{\mbox{\rm exp}}
\newcommand{\HH}{\mbox{$h$}}
\newcommand{\TT}{\mbox{$t$}}
\newcommand{\K}{\mbox{$k$}}
\newcommand{\euler}{\mbox{\normalsize{$\chi$}}(\s){\mbox{\normalsize{$/2$}}}}
\newcommand{\Char}{\mbox{\normalsize{$\chi$}}}

\newcommand{\MPL}{Mod. Phys. Lett. {\bf A}}
\newcommand{\NPB}{Nucl. Phys. {\bf B}}
\newcommand{\PRD}{Phys. Rev. {\bf D}}
\newcommand{\PLB}{Phys. Lett. {\bf B}}
\newcommand{\IJM}{Int. Jour. Mod. Phys. {\bf A}}
\newcommand{\A}{\mbox{$i$,$j$,$k$,$l$ all different}}
\newcommand{\B}{\mbox{$\delta^{ij}\varepsilon^{ik}\varepsilon^{il}
 \varepsilon^{kl}$}}
\newcommand{\C}{\mbox{$\delta^{ij}\delta^{jk}\varepsilon^{il}$}}
\newcommand{\D}{\mbox{$i=j=k=l$}}
\newcommand{\E}{\mbox{$\delta^{ij}\delta^{kl}\varepsilon^{ik}$}}

\begin{document}

\begin{flushright}
BRX-TH-376    
\end{flushright}
\begin{center}
{\Large{\bf Field Strength Correlators For Two Dimensional Yang-Mills Theories
\\ \vspace{.1in} Over Riemann Surfaces }}

Jo\~ao P. Nunes\footnote{supported by JNICT and FLAD (Portugal);
address after Sept.1, 1996, Dept. of Physics, Brown University, 
Providence RI 02912.}\\
and\\
Howard J. Schnitzer\footnote{supported in part by the DOE under grant
DE--FG02--92ER40706.}\\

Department of Physics\footnote{email address: NUNES,
SCHNITZER@BINAH.CC.BRANDEIS.EDU} \\
Brandeis University\\
Waltham, MA 02254

{\bf Abstract}
\end{center}

The path integral computation of field strength correlation functions
for two dimensional Yang-Mills theories over Riemann surfaces is studied.
The calculation is carried out by abelianization, which leads to correlators
that are topological. They are nontrivial as a result of the
topological obstructions to the abelianization.
It is shown in the large $N$ limit on the sphere that
the correlators undergo second order phase transitions at the critical point.
Our results are applied to a computation of contractible Wilson loops.

\section{Introduction}

In the last few years two-dimensional Yang-Mills theories have been studied
extensively. The partition function for the theory on \s, a Riemann surface
of genus $g$, was computed exactly in \cite{BR} (see also \cite{EW1}). More
recently Gross and Taylor have provided a string interpretation of the $1/N$
expansion of the partition function for the $SU(N)$ gauge group by enumerating
classes of maps from worldsheets to the target surface \s\ \cite{DG,JM,GT}.
The string description of the theory was later extended to the $SO(N)$ and
$Sp(2N)$ gauge groups
 in \cite{NRS}. Some progress has been made towards
finding
a string action for the 2d QCD string \cite{CMR,PH} as well.
However Douglas and Kazakov \cite{DK} have shown that the string
interpretation no longer holds for weak coupling when \s\ is a sphere, although
it does remain valid for strong coupling. On the sphere the large
$N$ theory
undergoes a phase transition at a critical value of $e^{2}(Area)=\pi^{2}$. The
study of this phase transition might also be relevant for four dimensional QCD,
which may or not exhibit a similar behaviour \cite{DK,GM}. It is
therefore important to understand those features of a gauge theory which give
rise to string behavior and to a phase-transition.

Two dimensional Yang-Mills theories have also been studied by means of
beautiful path integral methods. In \cite{EW2} the path integral was evaluated
by a generalized localization formula, while in \cite{BT1}, the path integral
for
gauge fields of fixed holonomy around the boundary of a disc was used as the
elementary building block from which the partition function and Wilson loops on
closed surfaces could be constructed.

In this paper we will study correlation functions of field strengths in 2d
non-abelian gauge
theories as an application of the abelianization technique for path integrals
developed by Blau and Thompson \cite{BT2,BT3,BT4}. We show that these
correlators exhibit the almost topological nature of the theory, and can be
expressed in terms of the higher order Casimir operators of the gauge group.
They can also be related to the generalized 2d QCD of \cite{DLS,RY} where
higher order Casimir operators appear in the partition function.
In section 2 we use abelianization to compute the partition function in the
presence of an external source. This allows us to calculate the correlators of
field strengths,
which is presented in section 3. One might expect the correlators of field
strengths to be trivial, since the Lagrangian with a source coupled to
$F_{\mu\nu}$ is quadratic in the fields. However, because of the topological
obstructions encountered in the abelianization, the correlators are highly
non-trivial. In section 4 we show that, on the
sphere, the regularized correlators of arbitrary numbers of field strength 
operators have second-order phase-transitions in the large $N$ limit. 
Section 5 shows how contractible Wilson loops can be computed from our
correlators, in agreement with the known formulae for Wilson loops. 
Our results suggest that, in the abelianization gauge, the usual Stokes'
 theorem
can be used to relate the Wilson loop to the field strength correlation
functions when the gauge group is $SU(2)$ [see equation (\ref{co5.1})]. 
In Appendix A we compute
correlators on the disc in a different gauge, while Appendix B is devoted to
some Lie algebra details.

The study of the correlators of the field strengths in the abelianization
gauge, 
enables us to calculate the master field for the regularized field strength 
$F_{\mu\nu}$ in
the large $N$ limit on the sphere, in a sequel to this paper 
\cite{JPNS}. 
Since, as we show in this paper, in this gauge these correlation
functions are essentially topological, $i.e.$ independent of position on the
Riemann surface, the master field will also have this property. This enables us
to construct the master field for $A_{\mu}$ itself on the sphere \cite{JPNS}.
In fact, the master field which exhibits the whole structure of the 
(unregulated) correlation functions, as is needed for the Wilson loop, can be 
obtained in a Hilbert space representation 
\cite{tese}.

One hopes to be able to calculate fermion correlation functions on the sphere
in the large $N$ limit, using the master field for
$A_{\mu}$. Thus, this paper serves as the first step in this program of
understanding the coupling of fermions to 2D Yang-Mills theory on Riemann
surfaces. To date this has only been acomplished on the plane \cite{GH}.
We are presently studying this application of our results.


\section{The Path Integral \ZJ}

In this section we consider the field strength of the $U(N)$ gauge theory on
 the compact surface of genus $g$,
\s, coupled to an external source $J(x)$ transforming in the adjoint of the
gauge group. The path integral describing this situation will be evaluated
using the elegant abelianization method of \cite{BT2,BT3}. The resulting
partition function \ZJ\ will then enable us to compute the electric field
correlators in the next section.

The partition function in question is,
\begin{equation}
  \ZJ=\int{\cal D}A_{\mu}\,\e[\frac{-1}{2e^{2}}\y\,\T(\xi^{2})+\y\,\T(J\xi)]
\label{2.1}
\end{equation}
where the scalar fields $\xi^{a}$$(x)$ are defined by 
$F_{\mu\nu}(x)=\xi(x)\sqrt{g(x)}\epsilon_{\mu\nu}$, where $\xi=T^{a}\xi^{a}$,
 with $T^{a}$ a generator of $U(N)$,
$d\mu=\sqrt{g(x)}d^{2}x$ a Riemannian measure on \s, 
$\epsilon_{\mu\nu}$ is the usual antisymmetric tensor with $\epsilon_{01}=1$,
and $\sqrt{g(x)}$ is the square root of the determinant of the metric on \s.

The starting point is the first order form of the path integral,
\begin{equation}
  \ZJ=\int{\cal D}\phi{\cal D}A_{\mu}\, \e [-\frac{e^{2}}{2}\y\,
\T(\p-iJ)^{2}-i\y\,
\T(\p\xi)]
\label{2.2}
\end{equation}
where \p\ is a Lie algebra($\cal G$) valued  scalar field also
 transforming in the adjoint under the gauge group. Performing the gaussian
 integration over \p\ gives back (\ref{2.1}). We decompose $\cal G$ into,
\begin{equation}
  \mbox{$\cal G$}=\TT\oplus\K
\label{2.3}
\end{equation}
where $\TT=\HH\oplus u(1)$ generates the usual maximal torus ($T$) of
$U(N)$, with \HH\ being the $su(N)$ Cartan subalgebra ($su(N)$ denotes the Lie
algebra of $SU(N)$, etc).

\setcounter{footnote}{0}
 The path integral (\ref{2.2}) then becomes the product of the integrals over
the $su(N)$ and $u(1)$ valued fields, where the two sectors will be related by a topological selection rule as explained ahead\footnote{This condition 
reflects the fact that we have $U(N)$ gauge group, not just $U(1)\times 
SU(N)$.}. 
We will evaluate each of these separately
beginning with the $U(1)$ sector. Bold type style quantities will be $su(N)$
valued and primed quantities will be $u(1)$ valued.
In order to account for the nontrivial $U(N)$ bundles over \s, we use the
topological condition for the $u(1)$ component of the gauge
curvature $\xi^{'}$,
\begin{equation}
  \y\,\xi^{'}=2 \pi\frac{m}{\sqrt{N}} , \;\;\;\;\;\; \mbox{              $m$ an integer.}
\label{2.4}
\end{equation}
It is convenient to decompose $m$ as $m=\tilde{m}\cdot N+p$ where $\tilde{m}$ 
is an integer and $p\in Z_{N}$. The topological selection rule connecting the $u(1)$ and $su(N)$ sectors will depend on $p$ and therefore we keep $p$ fixed for now. The sum over all possible values of $p$ will be performed when we examine the $su(N)$ sector.
To impose condition (\ref{2.4}) we insert the periodic delta function,
\begin{equation}
  \sum_{n} \e(i\frac{n}{\sqrt{N}} \y\,\xi^{'})\cdot\e(-2\pi i\frac{p}{N}n) 
\label{2.5}
\end{equation}
in the path integral. Let us consider the $u(1)$ piece first, using the
strategy developed in \cite{BT1}.
 The action is linear in $\xi^{'}$, and the
Nicolai map can be used to change variables of integration from $A_{\mu}^{'}$
to $\xi^{'}$,
and some scalar gauge fixing function $G(A_{\mu}^{'})$. The integration over
this second new variable, together with the jacobian from the change of
variables, will cancel against the Faddeev-Popov determinant. The integration
over $\xi^{'}$ now gives a $delta$ function that fixes the (constant)
value of $\phi^{'}$. The final contribution to (\ref{2.2}) from these $u(1)$
abelian fields is,
\begin{equation}
 \sum_{n} \e(\frac{e^{2}}{2}\y\,(J^{'})^{2} - \frac{e^{2}An^{2}}{2N} 
+i\frac{1}{\sqrt{N}}e^{2}n\y\,J^{'})\cdot\e(-2\pi i\frac{p}{N}n)
\label{2.6}
\end{equation}
where the source $J^{'}(x)$ just couples to $u(1)$ fields\footnote{We use
$J^{'}=\frac{\T (J)}{\sqrt{N}}$ and 
$\xi^{'}=\frac{\T (\xi)}{\sqrt{N}}$.}, and $A$ is the area
of \s. (We take ${\rm Tr}(T^{a}T^{b})=\delta^{ab}$ for the $U(N)$
generators). Notice that for $N=1$ there is no dependence on $p$ in 
(\ref{2.6}).
We will now do the remaining integrations over the $su(N)$ valued fields. Here
we will use the abelianization method of \cite{BT2,BT3}. One chooses the gauge
condition, which restricts $\phi$ to the Lie algebra of the maximal torus,
$i.e.$
\begin{equation}
 \PP^{\K} = 0
\label{2.7}
\end{equation}
in order to make use of the fact that the inner product (given by \T) on $\cal
G$ makes \HH\ and \K\ orthogonal.
First one must
cover \s\ by open sets where (\ref{2.7}) is valid, then in the intersections
of those open sets with each other, the \HH\ components of \PP\ (and also other
fields) will in general be related by nontrivial gauge transformations
preserving (\ref{2.7}) $i.e.$ by transformations which take values in $N(T)$,
the normalizer of the maximal torus $T$ in $G$. The remarkable fact that one
can choose those gauge transformations to lie in $T$ itself \cite{BT4}, so that
$\PP^{\HH}$ is globally defined, follows from the simple connectivity of 
$\cal G$. These $T$-valued gauge transformations then give rise to non-trivial $T$ bundles over \s. We refer to 
Blau and Thompson's \cite{BT4} for the detailed
explanations of this beautiful result. (Strictly speaking at this
point one must restrict the fields $\phi$ to take values in the regular
elements of
$\cal G$, but we'll see ahead that the non-regular valued fields are eliminated
by the Faddeev-Popov determinant \cite{BT4}).
Thus, the price to be paid
for demanding (\ref{2.7}) is the appearence of nontrivial $T$ bundle
topologies. These however can be easily incorporated in the path integral by
making use of the first Chern numbers \cite{BT2,BT3}.
We have then, for the integration over the $su(N)$ valued fields
\begin{equation}
 \frac{1}{|W|}\sum_{T bundles}\int{\cal D}A_{\mu}^{\HH}{\cal D}A_{\mu}^{\K}
{\cal D}\PP^{\HH}\, \mbox{det}[ \Delta_{W}(\PP^{\HH})]\e(-S(\J))
\label{2.8}
\end{equation}
where $|W|$ is the order of the Weyl group and det[$\Delta_{W}(\PP^{\HH})$] is
 the path integral analogue of the Weyl determinant in the Weyl integral
formula
 for Lie groups. This can also be interpreted as a Faddeev-Popov determinant
 coming from (\ref{2.7}), \cite{BT3}.
\begin{equation}
\mbox{det}[\Delta_{W}(\PP^{\HH})]=\mbox{det}[ad(\PP^{\HH})
|_{\Omega^{0}(\s,\K)}]
\label{2.9}
\end{equation}
where $\Omega^{0}(\s,\K)$ is the space of \K\ valued 0-forms on \s.

Using the orthogonality between \HH\ and \K\, the only
contribution from the integration over the \K\ components of the gauge
connection is easily seen to be given, up to a constant, by

\begin{equation}
 (\mbox{det}[ad(\PP^{\HH})|_{\Omega^{1}(\s,\K)}])^{-1/2}
\label{2.10}
\end{equation}
where $\Omega^{1}(\s,\K)$ is the space of \K\ valued 1-forms on \s. Notice the
similarity to (\ref{2.9}). In fact using the Hodge decomposition of
$\Omega^{1}(\s,\K)$ into orthogonal components, the contribution of (\ref{2.9})
and (\ref{2.10}) nearly cancels (see \cite{BT2} for a careful
evaluation of this ratio of determinants); that is
\begin{equation}
\frac{\mbox{det}[ad(\PP^{\HH})|_{\Omega^{0}(\s,\K)}]}
{(\mbox{det}[ad(\PP^{\HH})|_{\Omega^{1}(\s,\K)}])^{1/2}}=
\mbox{det}[ad(\PP^{\HH})|_{\K}]^{-b_{1}/2+b_{0}}=\mbox{det}[ad(\PP^{\HH})
|_{\K}]^{\euler}
\label{2.11}
\end{equation}
where $b_{0}=1$, $b_{1}=2g$ are the Betti numbers for \s\ and $\chi(\s)$ is the
Euler number of the surface. (We'll see that only
constant $\PP^{\HH}$ configurations contribute in the end).

At this stage we have reduced the integration over the $su(N)$-valued fields
to
\begin{equation}
\frac{1}{|W|}\sum_{T bundles}\int{\cal D}A_{\mu}^{\HH}{\cal D}\PP^{\HH}
\mbox{det}[ad(\PP^{\HH})|_{\K}]^{\euler}\;\{\e[-\frac{e^{2}}{2}\y\,\T(\PP^{\HH}-i\J)^{2}
-i\yy\,\T(\PP^{\HH}dA^{\HH})]\}
\label{2.12}
\end{equation}

The remaining $T$ valued gauge invariance, not eliminated by (\ref{2.7}), can
be fixed, and the corresponding Faddeev-Popov determinant eliminated by using
the
Nicolai map once more. In this way $dA^{\HH}$ is traded for $A_{\mu}^{\HH}$ as
a new variable in the path integral. It is now easy to express the summation
 over nontrivial $T$ bundles in terms of the first Chern numbers, as in
(\ref{2.4}) and (\ref{2.5}). We choose a set {$\alpha_{i}$}, $i=1,...,N-1$, of
simple roots for $su(N)$ and we let $(dA^{\HH})^{i}$ be the components of
$dA^{\HH}$ in the corresponding basis for \HH. 
Let us consider the topological sector with $u(1)$ charge $m$ as in 
(\ref{2.4}). If we write $m=\tilde{m}\cdot N+p$ with $p\in Z_{N}$ and 
$\tilde{m}$ an 
integer, then the topological conditions for $(dA^{\HH})^{j}$ are
\begin{equation}
 \yy\,(dA^{\HH})^{j}=2 \pi (m^{j}-\frac{pj}{N}), \;\;\;\;\; 
\mbox{        $m^{j}$ integers}
\label{2.13}
\end{equation}
This topological selection rule arises because the $U(1)$ and $SU(N)$ 
components of $U(N)$ are related since $U(N)\sim (U(1)\times SU(N))/Z_{N}$.
Inserted in the path integral, this condition will combine with the 
$\e(-2\pi i\frac{p}{N}n)$ factor in (\ref{2.6}) and will give origin to a 
summation 
over all pairs of representations of $U(1)$ and $SU(N)$ that can be combined
 into 
a representation of $U(N)$, namely such that $(n-R)\;{\rm mod}N\,=0$, where 
$R$ 
is 
the number of boxes of the Young tableau for the $SU(N)$ representation and $n$ labels the $U(1)$ representation as in (\ref{2.5}).
Therefore the summation over $T$ bundles and the remaining summation over $p$ 
can be done by inserting
\begin{equation}
 \sum_{\lambda\in \Lambda}\e[i\lambda (\yy\,dA^{\HH})]\;\cdot
\delta ((n-R(\lambda))\;{\rm mod}N)
\label{2.14}
\end{equation}
in the path integral. In the sum $\Lambda$ stands for all integer linear
combinations of the fundamental weights of $su(N)$. The remaining integrations
are immediate,
the integral over $dA^{\HH}$ producing $\delta(\PP^{\HH}+\lambda)$, so that
indeed only constant $\PP^{\HH}$ configurations contribute. One can now use the
classic formulas \cite{BT3,BD,FH},
\begin{equation}
 \mbox{det}[ad(\PP^{\HH})|_{\K}]\sim \prod_{\delta}(\beta,\PP^{\HH})
\label{2.15}
\end{equation}
and
\begin{eqnarray}
 \dim(\mu)=\prod_{\delta^{+}} \frac{(\beta,\mu+\rho)}{(\beta,\rho)}\\
 \nonumber \\
 C_{2}(\mu)=(\mu+\rho)\cdot (\mu+\rho)-\rho\cdot \rho
\label{2.16}
\end{eqnarray}
where $\dim(\mu)$ and $C_{2}(\mu)$ is the dimension and quadratic Casimir of
the irreducible representation of $SU(N)$ with highest weight $\mu$,
$\delta$ ($\delta^{+}$) is the set of all (positive) roots,
and $\rho$ is the half-sum of the positive roots.\footnote{The quantity
$(\mu+\rho)$ coincides with the variable $\sigma $ defined by Okubo \cite{O2},
his eqn.(15), in his study of third order Casimir operators.}

As remarked in \cite{BT2,BT3} the Weyl determinant in (\ref{2.15}) vanishes
whenever $\PP^{\HH}$ lies in the walls of a Weyl chamber. Thus the integration
should be restricted to regular valued $\PP^{\HH}$ fields, $i.e.$ those in the
interior of the Weyl chambers (see \cite{BT4} for a
detailed discussion). In the source free theory all $\PP^{\HH}$ can
then be rotated to the interior of the fundamental Weyl chamber, and the
$\frac{1}{|W|}$ factor gets cancelled. Every element of $\Lambda $
then corresponds to a unique element $(\mu+\rho)$ in the interior of the
fundamental Weyl chamber. However when the source $\J$ is not zero, one must
retain the average over the Weyl
group. In fact, that is why we have kept the harmless constant 
$\frac{1}{|W|}$ of
the Weyl integral formula in front of the path integral (\ref{2.12}). The
result for the $su(N)$ sector can then be written,
\begin{equation}
 \sum_{\mu}\dim(\mu)^{2-2g}\e[\frac{-e^{2}AC_{2}(\mu)}{2}+\frac{e^{2}}{2}\y\,\T(\J)^{2}]
\;\frac{1}{|W|}\sum_{\sigma\in W}\e[ie^{2}\y\,(\sigma(\mu+\rho),\J^{\HH})]
\label{2.17}
\end{equation}
where the sum goes over representations of $su(N)$ compatible with $u(1)$ 
charge $n$ as explained above.
In (\ref{2.17}) we have fixed the allowed renormalization terms
\cite{EW1,BT3} of the form $\exp(c_{0}e^{2}A)$ and $\exp(c_{1}(2-2g))$, to
give agreement with the usual normalization chosen for $\J=0$. We can now
combine the $u(1)$ and $su(N)$ pieces to give the total partition function
for $U(N)$,
\begin{eqnarray}
\ZJ=\sum_{l}\dim(l)^{2-2g}\exp(\frac{-e^{2}AC_{2}(l)}{2})
\;\{\e(\frac{e^{2}}{2}\y\,\T(J)^{2}) \nonumber \\ \times \frac{1}{|W|}\sum_{\sigma\in W}
\e(ie^{2}\y\,(\sigma(l+\rho),J^{\TT}))\}
\label{2.18}
\end{eqnarray}
where now $l$ runs over the highest weights of irreducible representations of
$U(N)$. (The $l^{i}$ are the row lengths of the $U(N)$ Young tableau for the
representation $l$.)
It should be emphasized that the last term in (\ref{2.18}) is a consequence of
the coupling of the source to the nontrivial $T$ topological sectors of the
gauge field.
Notice that for $U(N)$, the Weyl group $W$ is the symmetric group $S_{N}$.
This computation can also be trivially extended to the case of the generalized
2d QCD of \cite{DLS,RY}. In fact, since we have seen that only constant
configurations of $\PP^{\HH}$ contribute, our calculation also applies
when higher powers of $\PP$ are present.


\section{The Correlators}

In this section we will study the field strength correlation functions,
obtained from functional derivatives with respect to $J$ in \ZJ. Since the
average over the Weyl group
plays an essential role, it is convenient to use a basis for $\HH\oplus u(1)$
where the action of the Weyl group is as simple as possible. For $\K$ we will
use the usual
generators of the fundamental representation of $SU(N)$, and for the
generators of $T$ we will
use the diagonal matrices $E_{ii}$ $i=1,...,N$, where $E_{ii}$ has
only
one nonzero entry, $1$ on the $i^{th}$ position of the diagonal\footnote{In 
this basis one has that the $l^{i}$ are the row lengths of the Young tableau of the $U(N)$ representation $l$ and that $\rho^{i}=\frac{N+1}{2}-i$.}. 
We adopt the
normalization $\T(T^{a}T^{b})=\delta^{ab}$. The elements of
the Weyl group $S_{N}$ then just permute the entries of the diagonal matrices
in $\HH\oplus u(1)$. (The Weyl group acts as the identity on $u(1)$.)

As expected from the symmetry $\xi \rightarrow -\xi$
 of (\ref{2.1}) with $J=0$,
all odd point functions vanish. In our framework, the odd point functions
vanish since various products with an odd number of
factors of the form,
$$
   (l+\rho)^{\sigma_{i}}......(l+\rho)^{\sigma{j}}
$$
are averaged over $W$.
However, for each represention with highest
weight $l$ there exists a conjugate representation with highest weight
$\bar{l}$
such that $(\bar{l}+\rho)=-\bar{\sigma}(l+\rho)$, where
$\bar{\sigma}_{i}=N-i+1$.
Moreover the dimensions and quadratic
 casimirs of $l$ and $\bar{l}$ are
the same, so that the contribution of $l$ cancels that of $\bar{l}$ in the
sum over $l$ in the odd point functions, and therefore these vanish. 
(For self-conjugate
representations one will have $(l+\rho )=-\bar{\sigma}(l+\rho )$ so that the
average over the Weyl group also gives zero).

There are two terms which depend on $J$ in \ZJ. The term exp[$\y\,\T(J^{2})$]
gives rise to contact terms in the expectation values of the field strengths.
If one wants to consider 
products of fields at the same point one
needs to regulate these terms. This term is closely related to 
the arbitrary
renormalization term exp$(c_{0}e^{2}A)$ that appears in front of the
partition function. In fact by differentiating the $J=0$ partition function
with respect to $e^{2}$, we can conclude that the normalized correlator
\begin{equation}
 \langle\y\T\xi^{2}(x)\rangle =2e^{2}A\frac{d}{dA}F
\label{3.1}
\end{equation}

where $F=\mbox{\rm log }Z_{\s}$ is the free energy.\footnote{All correlators
will be normalized by dividing by the partition function. \\ 
That is
$\langle\xi^{a}(x)\cdots\xi^{b}(y)\rangle =\frac{1}{Z_{\Sigma_{g}}(0)}
\frac{\partial}{\partial J^{a}(x)}\cdots\frac{\partial}{\partial J^{b}(y)}
Z_{\Sigma_{g}}(J)|_{J=0}$}
Therefore in order to bring fields to the same point we must normalize
the contact term accordingly. This is done by adjusting the coefficient $c_{0}$
in the renormalization term.

For example, with this renormalization prescription the regularized 2-point 
function becomes
\begin{eqnarray}
\langle\xi^{a}(x)\xi^{b}(y)\rangle =
\frac{e^{4}}{Z_{\Sigma_{g}}}\sum_{l}\dim(l)^{2-2g}\e(\frac{-e^{2}AC_{2}(l)}{2})
  \makebox[1.8in]{} \nonumber  \\
\times [\frac{(\rho,\rho)\delta^{ab}}{N^{2}}\delta^{2}_{x,y}-\frac{1}{|W|}
\sum_{\sigma}(l+\rho)^{\sigma_{a}}(l+\rho)^{\sigma_{b}}]
\label{3.2}
\end{eqnarray}
where $\delta^{2}_{x,y}$ is $1$ if $x=y$, and zero otherwise. Notice the
$(l+\rho)$ terms are present {\em only when a and b are} $\HH\oplus u(1)$ 
 {\em indices}, $i.e.$ when the indices lie in the Cartan subalgebra. The 
average
over the Weyl group can be done explicitly with the result (see appendix B)
\begin{equation}
\frac{1}{|W|}
\sum_{\sigma}(l+\rho )^{\sigma_{a}}(l+\rho )^{\sigma_{b}}= 
 p^{ab}(l+\rho )^{2}+m^{ab}n^{2}
\label{weyl1}
\end{equation}
where
\begin{eqnarray}
p^{ab}= \left\{ \begin{array}{cc}
                 \frac{-1}{N(N-1)}  & \mbox{if $a\neq b$} \\
                 \frac{1}{N}        & \mbox{if $a=b$}
                \end{array}
        \right.
\nonumber
\makebox[1in]{and}
m^{ab}= \left\{ \begin{array}{cc}
                 \frac{1}{N(N-1)}  & \mbox{if $a\neq b$} \\
                 0                 & \mbox{if $a=b$}
                \end{array}
         \right.
\nonumber
\label{weyl2}
\end{eqnarray}
and
$$
n=\sum_{i}l^{i}=\mbox{total number of boxes in the tableau defined by $l$.}
$$

Thus we have the final result for the regularized 2-point function
\newpage
\begin{eqnarray}
\langle\xi^{a}(x)\xi^{b}(y)\rangle=\frac{e^{4}}{Z_{\Sigma_{g}}}
\sum_{l}\dim(l)^{2-2g}\e(\frac{-e^{2}AC_{2}(l)}{2}) \makebox[1.7in]{}
\nonumber \\
\times [\frac{(\rho,\rho)\delta^{ab}}{N^{2}}\delta^{2}_{x,y}-
(p^{ab}(l+\rho)^{2}+m^{ab}n^{2})]
\label{new1}
\end{eqnarray}
This is finite as $x\rightarrow y$.
{}From here (\ref{3.1}) follows as well.
Notice that the correlator is essentially topological in the gauge we have
chosen.

Similarly the regularized 4-point function is,
\begin{eqnarray}
\langle\xi^{i}(x)\xi^{j}(y)\xi^{k}(z)\xi^{l}(w)\rangle =
\frac{e^{8}}{Z_{\Sigma_{g}}}\sum_{l}\dim(l)^{2-2g}\e(\frac{-e^{2}AC_{2}(l)}{2})
\times
  \makebox[1.4in]{} \nonumber \\
\{\mbox{$[$}\frac{(\rho,\rho)^{2}}{(\dim G)^{2}}
(\delta^{ij}\delta^{2}_{x,y}
\delta^{kl}\delta^{2}_{z,w}+\mbox{2 permut.})\mbox{$]$}
-\frac{(\rho,\rho)}{\dim G}
\frac{1}{|W|}\sum_{\sigma}\mbox{$[$}\delta^{ij}\delta^{2}_{x,y}
(l+\rho)^{\sigma_{k}}
(l+\rho)^{\sigma_{l}}+\mbox{5 permut.}\mbox{$]$} \nonumber  \\
+\frac{1}{|W|}\sum_{\sigma}(l+\rho)^
{\sigma_{i}}(l+\rho)^{\sigma_{j}}(l+\rho)^{\sigma_{k}}(l+\rho)^{\sigma_{l}}\}
\makebox[2in]{}
\label{3.3}
\end{eqnarray}
(once again the $(l+\rho)$ terms are present {\em only for} $\HH\oplus u(1)$
{\em valued indices} in the Cartan subalgebra).
The average over $W$ gives (see appendix B),
\begin{eqnarray}
\frac{1}{|W|}\sum_{\sigma}(l+\rho)^{\sigma_{i}}(l+\rho)^{\sigma_{j}}
(l+\rho)^{\sigma_{k}}(l+\rho)^{\sigma_{l}}=
\makebox[1.6in]{}
\nonumber \\
a^{ijkl}\sum_{i}(l+\rho)_{i}^{4}+
b^{ijkl}[(l+\rho)^{2}]^{2}+c^{ijkl}n\sum_{i}(l+\rho)_{i}^{3}+
d^{ijkl}n^{2}(l+\rho)^{2}+e^{ijkl}n^{4}
\label{weyl3}
\end{eqnarray}
where the coefficents $a^{ijkl}$, etc can be found in equation (B.6).
Combining (\ref{3.3}) and (\ref{weyl3}) gives the regularized 4-point function
\begin{eqnarray}
\langle\xi^{i}(x)\xi^{j}(y)\xi^{k}(z)\xi^{l}(w)\rangle =
\frac{e^{8}}{Z_{\Sigma_{g}}}
\sum_{l}\dim(l)^{2-2g}\e(\frac{-e^{2}AC_{2}(l)}{2})\times
   \makebox[1.3in]{} \nonumber \\
\{\mbox{$[$}\frac{(\rho,\rho)^{2}}{(\dim G)^{2}}
(\delta^{ij}\delta^{2}_{x,y}
\delta^{kl}\delta^{2}_{z,w}+\mbox{2 permut.})\mbox{$]$}
-\frac{(\rho,\rho)}{\dim G}\mbox{$[$}\delta^{ij}\delta^{2}_{x,y}
(p^{kl}(l+\rho)^{2}+m^{kl}n^{2})+\mbox{5 permut.}\mbox{$]$}
\nonumber \\
+\mbox{$[$}a^{ijkl}\sum_{i}(l+\rho)_{i}^{4}+b^{ijkl}\mbox{$[$}
(l+\rho)^{2}\mbox{$]$}^{2}
+c^{ijkl}n\sum_{i}(l+\rho)^{3}_{i}+d^{ijkl}n^{2}(l+\rho)^{2}+e^{ijkl}n^{4}
\mbox{$]$}\}
\label{new2}
\end{eqnarray}
Once again we see that the correlator is topological.
Contracting with the group generators produces a gauge invariant quantity.
(Some useful identities for traces and higher order Casimirs can be found in
\cite{SO}). We have
\newpage
\begin{eqnarray}
\langle\T\xi^{4}(x)\rangle =
\frac{e^{8}}{Z_{\Sigma_{g}}}\sum_{l}\dim(l)^{2-2g}\e(\frac{-e^{2}AC_{2}(l)}{2})
\times  \makebox[1in]{} \nonumber
\\
\mbox{$[$}\frac{-(N^{2}-1)^{2}}{72N}(N^{2}+6N+\frac{1}{2})
-\frac{N^{2}-1}{3N}(N-1)C_{2}(l)-\frac{N^{2}-1}{3N}n^{2}+\frac{1}{4}C_{4}(l)
\mbox{$]$}
\label{3.4}
\end{eqnarray}
where 
$C_{4}(l)$ is a fourth order Casimir for $U(N)$ defined by
$C_{4}(l)=\sum_{i}[(l+\rho)^{i}]^{4}$.

The structure of the higher point functions is now clear. For the $2p$ point
function there's a Casimir invariant $C_{2p}$, and products of $(\rho,\rho)$
with lower order
casimirs of even order. The gauge invariant quantities
$\langle\T\xi^{2p}(x)\rangle$ have no dependence on $x$ at all, while the
non-invariant correlators are also topological and only depend on contact terms
in a trivial way. Thus, 
this gauge could be called a topological gauge. In fact
the theory is invariant under diffeomorphisms of \s\
that preserve the measure $d\mu$. Thus the $x$ dependence of gauge invariant
quantities could come only from
factors of $\sqrt{g(x)}$ which were already absorbed in the definition of
$\xi$ in (\ref{2.1}). Similarly, in the computation of Wilson loops only the
area enclosed by the loops was relevant (see for example \cite{BT1}). However,
notice that even the non-gauge invariant uncontracted correlators
(\ref{3.2}),(\ref{3.3}) above have only a trivial position dependence coming
from the contact term.
Therefore the gauge (\ref{2.7}) is particularly well suited to exhibit the
topological
nature of the theory, in the sense that expectation values of local non-gauge
invariant quantities in this gauge are also independent of position.


\section{The Phase Transition On $S^{2}$}

The partition function on the sphere is,
\begin{equation}
 Z=\sum_{R}\dim(R)^{2}\e(\frac{-AC_{2}(R)}{2N})
\label{4.1}
\end{equation}
where now $A=\lambda (Area)$, with $\lambda =e^{2}N$ held fixed when
$N \rightarrow \infty$.
Although the exponential decreases rapidly for large representations, for $A$
sufficiently small the series diverges due to the dimension term raised to a
positive power.
In fact, Douglas and Kazakov \cite{DK} have shown that in 
the large $N$ limit
the theory undergoes a third-order phase transition at $A_{cr}=\pi^{2}$. This
phase
transition was also explained in \cite{GM} as a consequence of the presence of
instantons (classical solutions of the field equations). When $A$ approaches
$\pi^{2}$, the contribution of the instanton configurations to the
partition function becomes dominant, and the phase transition is induced. The
relation to the string interpretation was also explored in \cite{WT,CNS}. It
has been shown that a 
similar phase transition takes place for ${\rm YM}_{2}$
on the cylinder \cite{CDMP}, on the projective plane $RP^{2}$, and for
$Sp(N)$ and $SO(N)$ on the sphere as well, \cite{CS}.
Further, the generalized 2d Yang-Mills theory possesses a rich phase structure,
with phase transitions possible for $g>1$, and for particular regions of the
coupling constants space \cite{RY}. The Wilson loops in the large $N$ limit
have been studied in both the weak and strong coupling phases
\cite{DK,BOU,R,DAK,CS}.

Here we will study the phase transition on the sphere for the electric field
correlators of ${\rm YM}_{2}$ in the large $N$ limit. Let
us briefly review the methods and known results. The irreducible
representations of $U(N)$ are labelled by the Young tableau row lengths
$l^{1}\geq l^{2}\geq ....\geq l^{N}$. Also,
\begin{equation}
(l+\rho)^{i}=\frac{N+1}{2}-i+l^{i}  \;\;\;\;  \mbox{, $i=1...N$}
\label{4.2}
\end{equation}
In the large $N$ limit, these variables can be replaced by continum variables
\cite{RU},
\begin{eqnarray}
x=\frac{i}{N} \makebox[1in]{with} 0\leq x\leq 1  \nonumber \\
l(x)=\frac{l^{i}}{N} \makebox[1in]{and} \sum_{i}=N\int_{0}^{1}dx
\label{4.3}
\end{eqnarray}
It is then useful to define,
\begin{equation}
h(x)=-l(x)+x-\frac{1}{2}
\label{4.4}
\end{equation}
When one expresses the partition function (\ref{4.1}) in terms of these
continum variables, the sum over representations becomes a quantum mechanical
functional integral for,
\begin{equation}
S_{eff}(h)=-\int_{0}^{1}dx
\int_{0}^{1}dy\;\mbox{\rm
log}[h(x)-h(y)]+\frac{A}{2}\int_{0}^{1}dxh^{2}(x)-\frac{A}{24}
\label{4.5}
\end{equation}
with
\begin{equation}
Z=\int{\cal D}h\,\e(-N^{2}S_{eff}(h))
\label{4.6}
\end{equation}
Since $N$ is large, this can be computed by the saddle-point approximation
\cite{RU,DK}. Define the density of boxes in a Young tableau by
\begin{equation}
u(h)=\frac{\partial x(h)}{\partial h} \mbox{, where $u(h)\leq 1$}
\label{4.7}
\end{equation}
The saddle point equation is
\begin{equation}
{\rm P}\int ds\frac{u(s)}{h-s}=\frac{A}{2}h
\label{4.8}
\end{equation}
For $A< A_{cr}=\pi^{2}$, the solution to this equation is given by the
semicircle law \cite{RU},
\begin{equation}
u_{weak}(h)=\frac{A}{2\pi}\sqrt{\frac{4}{A}-h^{2}} \mbox{, with $0\leq h\leq
\frac{2}{\sqrt{A}}$}
\label{4.9}
\end{equation}
However for $A>A_{cr}=\pi^{2}$ the above solution violates the condition
$u(h)\leq 1$ in (\ref{4.7}), and a different solution must be sought \cite{DK}.
The ansatz
\begin{eqnarray}
u_{strong}(h)=\left\{ \begin{array}{cc} 1 & \mbox{if $-b\leq h\leq b$} \\
\tilde{u}(h)
& \mbox{otherwise} \end{array}\right.
\label{4.10}
\end{eqnarray}
gives the solution
\begin{eqnarray}
 f(h)=\int ds\frac{\tilde{u}(s)}{s-h}
 \makebox[2.2in]{} \nonumber \\
\nonumber \\
=-h\frac{A}{2}-{\rm log}\frac{h-b}{h+b}+
\sqrt{(a^{2}-h^{2})(b^{2}-h^{2})}\int_{-b}^{b}
\frac{ds}{(h-s)\sqrt{(a^{2}-s^{2})(b^{2}-s^{2})}}
\label{4.11}
\end{eqnarray}
Following \cite{DK} we will now present the behaviour of the free energy,
$F=\frac{1}{N^{2}}{\rm log}Z$, at the critical point. This will enable us to
review the method and notation needed to describe the phase transition of the
correlators. We have,
\begin{equation}
F^{'}=\frac{dF}{dA}=-\frac{\partial S_{eff}(h)}{\partial
A}=-\frac{1}{2}\int_{0}^{1}dxh^{2}+\frac{1}{24}
\label{4.12}
\end{equation}
where one uses the fact that $u(h)$ is an even function both in the strong and
weak coupling solutions. When $A<\pi^{2}$ from (\ref{4.9}),
\begin{equation}
F^{'}(A)=\frac{1}{24}-\frac{1}{2A}
\label{4.13}
\end{equation}

For the strong coupling phase we need to use the large $h$ expansion of
(\ref{4.11}). The $O(h)$ and $O(h^{-1})$ terms give,
\begin{equation}
a=\frac{4K}{A}
\label{4.14}
\end{equation}
and
\begin{equation}
 A=8EK-4k^{'2}K
\label{4.15}
\end{equation}
where $k=b/a$ is the modulus of the elliptic integrals, $k{'}=\sqrt{1-k^{2}}$
is the complementary modulus and $K=K(b/a)$, $E=E(b/a)$ are the complete
elliptic integrals of the first and 
second kind respectively (see for example
\cite{BF}). To explore the behaviour near the critical point, it is useful to
express these quantities in terms of theta functions on a torus of complex
modulus $\tau$ \cite{DK,BF}. One sets $q=\exp{i\pi\tau}$ and
$\theta_{i}=\theta_{i}(0|\tau) \mbox{,       $i=0,1,2,3$}$. When we approach
the critical point, $b\rightarrow 0$ and $q\rightarrow 0$. We then define
$\delta=A-A_{cr}=A-\pi^{2}$. Expanding (\ref{4.15}) in $q$ gives,
\begin{equation}
 A=\pi^{2}(1+8
q-8q^{2}+32q^{3}+\cdots)
\label{4.16}
\end{equation}
and inverting for $\delta$ we find,
\begin{equation}
q=\frac{\delta}{8\pi^{2}}(1+\frac{\delta}{8\pi^{2}}+\cdots )
\label{4.17}
\end{equation}
{}From the weak and strong coupling expansions of (\ref{4.13}) one gets,
\begin{equation}
F^{'}_{strong}-F^{'}_{weak}=-\frac{\delta^{2}}{\pi^{6}}+\cdots
=-\frac{1}{\pi^{2}}\frac{(A-A_{cr})^{2}}{\pi^{4}}+\cdots
\label{4.18}
\end{equation}
which is the third-order phase-transition for the free energy of 
\cite{DK}.

{}From (\ref{3.1}), section 3, in the large $N$ limit\footnote{In this section
$z$ will denote a point on
$S^{2}$ and $x$ will always be the continuous index $x=\frac{i}{N}$.}
(note that here $F=\frac{1}{N^{2}}\log{Z}$),
\begin{equation}
\frac{1}{N}\langle\T\xi^{2}(z)\rangle=2\lambda\frac{dF}{dA}=2\lambda F{'}(A)
\label{4.19}
\end{equation}
and therefore the gauge-invariant two-point function behaves as $F^{'}(A)$ at
the critical point, with a second-order phase-transition.

The gauge invariant four-point function was given in (\ref{3.4}). In the large
 $N$
limit, only the critical representation survives at the saddle point.
Moreover, since the critical representation is self-conjugate, $h(x)$ is odd
about $x=\frac{1}{2}$, from (\ref{4.4}) the corresponding $u(1)$ charge
$n=\int_{0}^{1}l(x)$ vanishes.
Notice that,
$$
(l+\rho)(x)=\frac{1}{N}(l+\rho)^{i}=-h(x)
$$
then
\begin{equation}
\frac{1}{N}\langle\T\xi^{4}(z)\rangle=\lambda^{4}\{ \frac{-1}{72}-\frac{1}{3}
[\int_{0}^{1}dxh^{2}(x)-\frac{1}{12}]+\frac{1}{4}\int_{0}^{1}dxh^{4}(x)\}
\label{4.20}
\end{equation}
In the weak phase we can use (\ref{4.9}) to obtain,
\begin{equation}
\int_{0}^{1}dxh_{weak}^{4}(x)=\frac{2}{A^{2}}
\label{4.21}
\end{equation}
{}From the $O(h^{-5})$ term in (\ref{4.11}) in the strong coupling phase we
have,
\begin{equation}
\int_{0}^{1}dxh_{strong}^{4}(x)=\frac{4\pi^{6}}{5A^{5}}(\theta_{3}^{12}-
\theta_{2}^{4}\theta_{3}^{8}-\theta_{2}^{8}\theta_{3}^{4}+\theta_{2}^{12})+
\frac{6\pi^{4}}{5A^{4}}(\theta_{3}^{8}+\frac{2}{3}\theta_{3}^{4}\theta_{2}^{4}
+\theta_{2}^{8})
\label{4.22}
\end{equation}
Expanding for small $q$ and using (\ref{4.17}) we find,
\begin{equation}
\int_{0}^{1}dxh_{strong}^{4}(x)-\int_{0}^{1}dxh_{weak}^{4}(x)=\frac{4}{\pi^{4}}
\frac{(A-A_{cr})^{2}}{\pi^{4}}+\cdots
\label{4.23}
\end{equation}
so that the phase-transition for the four-point function is also second-order.

Similarly we can use the $O(h^{-7})$ term in (\ref{4.11}) to show that,
\begin{eqnarray}
-\int_{0}^{1}dxh_{strong}^{6}(x)=
\makebox[2.2in]{}
\nonumber \\
\frac{(2\pi)^{8}}{2A^{7}}[
\frac{151}{2688}(\theta_{3}^{16}+\theta_{2}^{16})-\frac{19}{672}
(\theta_{3}^{12}\theta_{2}^{4}+\theta_{2}^{12}\theta_{3}^{4})-\frac{25}{448}
\theta_{3}^{8}\theta_{2}^{8}-\frac{5}{48}\theta_{3}^{4}\theta_{0}^{8}\theta_{2}
^{4}-\frac{7}{96}\theta_{0}^{8}(\theta_{3}^{8}+\theta_{2}^{8})] \nonumber \\
+\frac{(2\pi)^{6}}{A^{6}}[-\frac{4}{21}(\theta_{3}^{12}+\theta_{2}^{12})+
\frac{5}{42}(\theta_{3}^{8}\theta_{2}^{4}+\theta_{2}^{8}\theta_{3}^{4})+
\frac{7}{48}\theta_{0}^{8}(\theta_{3}^{4}+\theta_{2}^{4})]
\makebox[1in]{}
\label{new}
\end{eqnarray}
so that,
\begin{equation}
\int_{0}^{1}dxh_{strong}^{6}(x)-\int_{0}^{1}dxh_{weak}^{6}(x)=
\frac{12}{\pi^{6}}\frac{(A-A_{cr})^{2}}{\pi^{4}}+\cdots
\label{4.24}
\end{equation}
Therefore the six-point function also has a second-order phase
transition at the critical point. (In fact it has been shown that
$$
\int_{0}^{1}dxh_{strong}^{2n}(x)-\int_{0}^{1}dxh_{weak}^{2n}(x)
$$
has a second-order transition for any $n$, so that all higher-point correlators
of the field strength have second-order phase-transitions as well
\cite{inprep}).

The master field for the field strength and gauge potential, on the sphere in
the large $N$ limit, based on the above results, which reproduces the regulated
correlators is presented in \cite{JPNS}. The unregulated correlators, that is
including non-regularized contact terms, can also be described by an explicit 
master field representation \cite{tese}.

\section{Application to Wilson Loops}

One of the motivations for studying field strength correlators is to 
understand 
the couplings of matter to 2D Yang-Mills theory on Riemann surfaces. To
illustrate this direction of research, we compute expectation values for
contractible Wilson loops using a version of Stokes' theorem, equation
(\ref{co5.1}), and our correlators and compare these calculations with 
known exact results.
Although we do not have a complete proof of the validity of this version of the
theorem, our results are so compelling, that we include them so as to stimulate
further research. To be specific,
we compute contractible Wilson loops on $\Sigma_{g}$ for $U(1)$ and $SU(2)$
gauge theories, using Stokes' theorem to relate the holonomy of the gauge field
in the representation $r$
around a closed loop to the exponential of the integral of the gauge curvature
over the interior of the loop. That is we consider
\begin{equation}
\langle W_{r}(\gamma)\rangle=\langle \T\;{\cal P}\exp(i\oint_{\gamma} A)
\rangle=\langle\T\exp(i\int_{D} F)\rangle
\label{co5.1}
\end{equation}
where we will take a contractible,
non-self-intersecting loop $\gamma=\partial D$. 
Obviously (\ref{co5.1}) is valid for a $U(1)$ gauge theory. However, motivated
by the abelianization of the partition function, we attempt to use
(\ref{co5.1}) for the $SU(2)$ gauge theory as well. As we describe below, we 
have
computed $\langle W(\gamma)\rangle$ for $SU(2)$ up to $O(\Delta^{4})$, where
$\Delta$ is the area enclosed by the loop $\gamma$, and verified that the
Stokes' theorem (\ref{co5.1}) is valid for this case as well, in the 
Blau-Thompson
abelianization gauge. It should be emphasized that (\ref{co5.1}) appears to
work for non-abelian theories, with gauge group $SU(2)$, computed in the 
abelianization gauge, as a result
of explicit calculations. Unfortunately, we cannot provide a proof of this
remarkable result, so that the application of (\ref{co5.1}) to $SU(2)$ non-abelian theory in this gauge remains a conjecture. We note that for non-abelian
Yang-Mills theories in general gauges, one must use a non-abelian version of 
the
Stokes' theorem, where one of the variables in the surface integral is ordered
and the gauge curvature appears conjugated by holonomies, so that gauge
invariance is preserved \cite{Stokes}. By contrast, in (\ref{co5.1}) there is 
no ordering in the integration over the disc and $F$ is not 
conjugated by holonomies. Nevertheless, it apparently computes the Wilson loop
in the abelianization gauge.

The expectation values of Wilson loops in two dimensional Yang-Mills theory are
known; see for example \cite{BT1}. For a loop $\gamma=\partial D$ on \s, in the
representation $r$ we have the exact result
\begin{equation}
\langle W_{r}(\gamma)\rangle =\sum_{l}
{\rm dim}(l)^{1-2g}\exp(\frac{-e^{2}AC_{2}(l)}{2})
\sum_{\lambda\in l\otimes r}{\rm dim}(\lambda)\exp[\frac{e^{2}\Delta
(C_{2}(l)-C_{2}(\lambda))}{2}] 
\label{co5.2}
\end{equation}
where $\Delta$ is the area of the disc $D$ enclosed by the Wilson loop. 
The exponential in
(\ref{co5.2}) is expanded in powers of $\Delta$, which then expresses 
$\langle W_{r}(\gamma)\rangle$ as a power series in $\Delta$.
On the other hand, one can expand the right-hand side of (\ref{co5.1}) as
\begin{equation}
\langle W_{r}(\gamma)\rangle=\T(1)-\frac{1}{2}
\langle\T \int_{D} F(x)\int_{D}F(y)
\rangle
+\frac{1}{4!}\langle\T (\int_{D} F)^4\rangle +\cdots
\label{co5.3}
\end{equation}

where we have used our result that odd correlators of field strengths vanish.
Comparing the expansion of (\ref{co5.2}) in powers of $\Delta$ with
(\ref{co5.3}) seems to present a paradox, as (\ref{co5.2}) contains odd powers
of $\Delta$, while (\ref{co5.3}) only involves an even number of integrations
on the disc $D$. In fact, there is a very elegant resolution of this issue when
one uses the (unregulated) correlation functions of section 3, as these
correlators have contributions from the contact terms, obtained by means of
functional derivatives with respect to the source from the ${\rm Tr}(J^{2})$
term in (\ref{2.18}). We will refer to contributions from this term, as the
contact terms of the correlators, and from the linear term in $J$ in 
(\ref{2.18}) as the topological terms.

For example, the contact term ($\sim \sqrt{g(x)}\sqrt{g(y)}\delta^{2}(x-y)$) 
in the two 
point function will produce a linear term in $\Delta$ in (\ref{co5.3})
when integrated over the disc.
This is also the case for the Wilson loop on the plane. See for example 
Brali\'c in \cite{Stokes}. 
In general each contact term and each topological term in a field strength
correlator produces a term linear in $\Delta$ when inserted in (\ref{co5.3}).

We now turn to the explicit verification of (\ref{co5.3}), correct to
$O(\Delta^{4})$, with $U(1)$ and $SU(2)$ considered separately for clarity. 

\subsection{$U(1)$ gauge theory}

The Wilson loop in an arbitrary representation $r$ of $U(1)$ can be written as
\begin{equation}
\langle
W_{r}(\gamma)\rangle=\sum_{n}\exp(\frac{-e^{2}An^{2}}{2})
\exp(\frac{-e^{2}\Delta (r^{2}+2nr)}{2})
\label{co5.4}
\end{equation}
where $n$ runs over all integers. 

The expansion of (\ref{co5.4}) in powers of $\Delta$ yields
\newline
\begin{eqnarray}
\langle W_{r}(\gamma)\rangle
=\sum_{n=-\infty}^{\infty}\exp(-\frac{e^{2}An^{2}}{2})\cdot 
\makebox[3.3in]{} \nonumber \\
\cdot
\{1-\frac{e^{2}}{2}r^{2}\Delta +\frac{1}{2!}\frac{e^{2}}{4}(r^{4}+4n^{2}r^{2})
\Delta^{2}-\frac{1}{3!}\frac{e^
{6}}{8}(r^{6}+12n^{2}r^{4})\Delta^{3}+
\makebox[1in]{} \nonumber \\
+\frac{1}{4!}\frac{e^{8}}{16}(r^{8}+24n^{2}r^{6}+16n^{4}r^{4})\Delta^{4}+
\cdots \}\makebox[0in]{}  
\label{co5.5a}\\
\equiv \langle 1\rangle -\frac{e^{2}}{2}\Delta\langle r^{2}\rangle
+\frac{e^{2}}{8}\Delta^{2}\langle r^{4}+4n^{2}r^{2}\rangle
-\frac{e^{6}}{48}\Delta^{3}\langle r^{6}+12n^{2}r^{4}\rangle + 
\makebox[1in]{} \nonumber \\
+\frac{1}{4!}\frac{e^{8}}{16}\Delta^{4}\langle r^{8}+24n^{2}r^{6}+
16n^{4}r^{4}\rangle +\cdots   
\makebox[0in]{}
\label{co5.5b}
\end{eqnarray}
where the expectation values in (\ref{co5.5b}) are defined in the obvious way
from (\ref{co5.5a}).

Our objective is to compute the right-hand side of (\ref{co5.3}) and to
compare it with (\ref{co5.5b}). The relevant generating functional of 
correlation
functions is (\ref{2.6}), specialized to $N=1$, and with each $F$ multiplied by
$r$ in accord with our normalization conventions. Let us sketch the
explicit comparisons, using (\ref{co5.3}) a
nd our unregulated correlators.
For convenience we write $F_{c}$ or $F_{t}$ for the part of the field strength
correlators coming from the contact terms or topological terms in the
generating functional. We indicate these terms in the expectation values
schematically, as these usually correspond to a number of permutations of
contact and topological terms. We stress that $F_{c}$ and $F_{t}$ have only
symbolic meaning, as labels of terms in the correlation functions.

$\underline{O(\Delta)}$:
\begin{equation}
-\frac{1}{2}\int_{D}dx\int_{D}dy\;\langle F_{c}(x)F_{c}(y)\rangle =
\langle r^{2}\rangle (-\frac{1}{2})\int_{D}d\mu (x)\int_{D}
d\mu (y)\;
e^{2}\delta^{2}(x-y)=-\frac{e^{2}}{2}\langle r^{2}\rangle\Delta
\label{co5.6}
\end{equation}
which agrees with (\ref{co5.5b}).

$\underline{O(\Delta^{2})}$:
\begin{equation}
-\frac{1}{2}\int_{D}dx\int_{D}dy\;\langle F_{t}(x)F_{t}(y)\rangle =
(ie^{2})^{2}r^{2}(-\frac{1}{2})\langle n^{2}\rangle\Delta^{2}
\label{co5.7a}
\end{equation}
and
\begin{equation}
\frac{1}{4!}(\int_{D})^{4}\langle F_{c}^{4}\rangle =
\frac{1}{4!}\langle r^{4}\rangle e^{4}(\int_{D})^{4}
[\delta^{2}(x-y)\delta^{2}(z-w)+ 2\;{\rm permut.}]=
\frac{1}{8}e^{4}\langle r^{4}\rangle\Delta^{2}
\label{co5.7b}
\end{equation}
The sum of (\ref{co5.7a}) and (\ref{co5.7b}) agrees with the $O(\Delta^{2})$
term in (\ref{co5.5b}). Further, the two terms in (\ref{co5.5b}) of order
$O(\Delta^{2})$ originate from (\ref{co5.7b}) and (\ref{co5.7a})
respectively.

$\underline{O(\Delta^{3})}$:
\begin{equation}
\frac{1}{4!}(\int_{D})^{4}\langle F_{c}^{2}F_{t}^{2}\rangle =
\frac{1}{4!}r^{4}(-e^{4})\langle n^{2}\rangle (\int_{D})^{4}\;
e^{2}[\delta^{2}(x-y)+5\;{\rm permut.}]=
-\frac{e^{6}}{4}r^{4}\langle n^{2}\rangle\Delta^{3}
\label{co5.8}
\end{equation}
and
\begin{equation}
-\frac{1}{6!}(\int_{D})^{6}\langle F_{c}^{6}\rangle =
-\frac{1}{6!}\langle r^{6}\rangle (\int_{D})^{6}\;e^{6}
[\delta^{2}(x-y)\delta^{
2}(z-w)\delta^{2}(v-u)+14\;{\rm permut.}]=
-\frac{e^{6}}{48}\langle r^{6}\rangle\Delta^{3}
\label{co5.9}
\end{equation}
Again, the agreement with (\ref{co5.5b}) is evident.

$\underline{O(\Delta^{4})}$:
\begin{equation}
\frac{1}{4!}(\int_{D})^{4}\langle F_{t}^{4}\rangle =
\frac{e^{8}}{4!}r^{4}\langle n^{4}\rangle\Delta^{4}
\label{co5.10a}
\end{equation}
while
\begin{equation}
-\frac{1}{6!}(\int_{D})^{6}\langle F_{c}^{4}F_{t}^{2}\rangle =
\frac{e^{8}}{16}r^{6}\langle n^{2}\rangle\Delta^{4}
\label{co5.10b}
\end{equation}
and
\begin{equation}
\frac{1}{8!}(\int_{D})^{8}\langle F_{c}^{8}\rangle =
\frac{e^{8}}{4!16}\langle r^{8}\rangle\Delta^{4}
\label{co5.10c}
\end{equation}
Thus, (\ref{co5.10a}) to (\ref{co5.10c}) agree with the $O(\Delta^{4})$ term of
(\ref{co5.5b}).

Therefore, we have shown how to obtain the expansion of the Wilson loop 
expectation value in powers of the area, for a
contractible non-self-intersecting loop, directly from the field strength 
correlation functions.

We note that each term in (\ref{co5.5b}) comes from a distinct combination
 of contact
and topological terms in the expectation values. This sets the stage for a
similar calculation for $SU(2)$.

\subsection{SU(2): Wilson Loop in the Fundamental Representation}

The Wilson loop expectation value of a gauge field in the fundamental
representation for a contractible non self-intersecting loop is
\begin{equation}
\langle W_{f}(\gamma)\rangle =\sum_{j=0}^{\infty}\dim(j)^{1-2g}
\exp(\frac{-e^{2}AC_{2
}(j)}{2})\sum_{\lambda\in j\otimes f}\dim(\lambda)
\exp(\frac{e^{2}\Delta(C_{2}(j)-C_{2}(\lambda))}{2})
\label{co5.11}
\end{equation}
where $\dim(j)=j+1$, $C_{2}(j)=\frac{1}{2}j(j+2)$ and $j\otimes f=(j-1)\oplus
(j+1)$ for $j\geq 1$. The expansion of (\ref{co5.11}) in powers of $\Delta$
gives, after some elementary but tedious algebra
\begin{eqnarray}
\langle W_{f}(\gamma)\rangle =
\sum_{j} \dim (j)^{2-2g}\exp(-\frac{e^{2}AC_{2}(j)}{2})\cdot    
\makebox[3in]{} \nonumber \\
\cdot\{\T(1)-\frac{3}{2}e^{2}{\Delta}+\frac{e^{4}}{16}[4(j+1)^{2}+5]\Delta^{2}
-\frac{1}{4}\frac{e^{6}}{48}[20(j+1)^{2}+7]\Delta^{3}+
\makebox[1in]{} \nonumber \\
+\frac{1}{4!8}\frac{e^{8}}{16}[16(j+1)^{4}+56(j+1)^{2}+9]\Delta^{4}+\cdots\}
\makebox[0in]{}   \label{co5.12} \\ 
=\langle 2\rangle -\frac{3}{2}e^{2}\Delta\langle 1\rangle
+\frac{e^{4}}{16}\Delta^{2}\langle 4(j+1)^{2}+5\rangle
-\frac{1}{4}\frac{e^{6}}{48}\Delta^{3}\langle 20(j+1)^{2}+7\rangle +
\makebox[1in]{} \nonumber \\
+\frac{1}{4!}\frac{e^{8}}{128}
\Delta^{4}\langle 16(j+1)^{4}+56(j+1)^{2}
+9\rangle+\cdots  \makebox[0in]{}
\label{co5.13}
\end{eqnarray}

The generating functional for the correlators of field strengths in the
fundamental representation is given by (\ref{2.17}). For $SU(2)$ it is
convenient to carry out the Weyl sum in (\ref{2.17}) directly, with the result
\begin{equation}
Z_{\s}(J)=\sum_{j=0}^{\infty}\dim(j)^{2-2g}\exp(\frac{-e^{2}AC_{2}(j)}{2})
\exp(\frac{e^{2}}{2}\y \T(J^{2}))\cos(\frac{e^{2}}{\sqrt{2}}(j+1)\y J^{'})
\label{co5.14}
\end{equation}

where $\frac{1}{\sqrt{2}}J^{'}\sigma^{3}$ is the diagonal component of 
$J$, with our choice of normalization. Let us compute the right-hand side of
(\ref{co5.3}) using the correlators generated by (\ref{co5.14}).

$\underline{O(\Delta)}$:
\begin{equation}
-\frac{1}{2}\int_{D}dx\int_{D}dy \langle \T (F_{c}(x)F_{c}(y))\rangle =
e^{2}\langle\delta^{ab}\T (T^{a}T^{b})\rangle (-\frac{1}{2})\int_{D}d\mu (x)
\int_{D}d\mu (y)\delta^{2}(x-y)=-\frac{3}{2}e^{2}\langle 1\rangle\Delta
\label{co5.15}
\end{equation}

which agrees with (\ref{co5.13}).

$\underline{O(\Delta^{2})}$:
$$
(\int_{D})^{4}\langle \T (F_{c}^{4})\rangle
\makebox[1.8in]{is obtained from}
\makebox[2.7in]
{$e^{4}
[\delta^{ij}\delta^{kl}\delta^{2}(x-y)\delta^{2}(z-w)+2\;{\rm permut.}]$ }
$$
which must be contracted with traces of the generators with structure
$\T (T^{i}T^{j}T^{i}T^{j})$ and $(\T (T^{i}T^{i}))^{2}$ to obtain
\begin{equation}
\frac{1}{4!}(\int_{D})^{4}\langle \T(F_{c}^{4})\rangle =
\frac{5}{16}e^{4}\langle 1\rangle\Delta^{2}
\label{co5.16a}
\end{equation}
Similarly
\begin{equation}
-\frac{1}{2}(\int_{D})^{2}\langle \T (F_{t}^{2})\rangle =
\frac{e^{4}}{4}\langle (j+1)^{2}\rangle 
\T [\frac{1}{\sqrt{2}}\sigma^{3}\frac{1}{\sqrt{2}}\sigma^{3}]\Delta^{2}
=\frac{e^{4}}{4}\langle (j+1)^{2}\rangle\Delta^{2}  
\label{co5.16b}
\end{equation}
Equations (\ref{co5.16a}) and (\ref{co5.16b}) agree with the $O(\Delta^{2})$
term in (\ref{co5.13}). (Note that only generators in the Lie algebra of the
maximal torus contribute to the topological terms.)

$\underline{O(\Delta^{3})}$:

The calculations become increasingly lengthy, so we only summarize the results
\begin{equation}
\frac{1}{4!}(\int_{D})^{4}\langle \T (F_{c}^{2}F_{t}^{2})\rangle =
-e^{6}\frac{5}{4!2}\Delta^{3}\langle (j+1)^{2}\rangle
\label{co5.17}
\end{equation}
and
\begin{equation}
-\frac{1}{6!}(\int_{D})^{6}\langle \T (F_{c}^{6})\rangle =
-\frac{8}{6!}\frac{105}{32}e^{6}\langle 1 \rangle\Delta^{3}
\label{co5.18}
\end{equation}
Equations (\ref{co5.17}) and (\ref{co5.18}) agree with the $O(\Delta^{3})$ 
term of (\ref{co5.13}).

$\underline{O(\Delta^{4})}$:
\begin{equation}
\frac{1}{4!}(\int_{D})^{4}\langle \T (F_{t}^{4})\rangle =\frac{e^{8}}{4!8}
\Delta^{4}\langle (j+1)^{4}\rangle
\label{co5.19}
\end{equation}
while
\begin{equation}
-\frac{1}{6!}(\int_{D})^{6}\langle \T (F_{c}^{4}F_{t}^{2})\rangle =
\frac{105}{6!8}e^{8}\Delta^{4}\langle (j+1)^{2}\rangle
\label{co5.20}
\end{equation}
and
\begin{equation}
\frac{1}{8!}(\int_{D})^{8}\langle \T (F_{c}^{8})\rangle =
\frac{9}{4!128}e^{8}\langle 1\rangle\Delta^{4}
\label{co5.21}
\end{equation}

The computation of (\ref{co5.19}) to (\ref{co5.21}) is very lengthy, but
nevertheless agrees with (\ref{co5.13}).
Note that in the expectation values $F_{t}^{2}$ gives
$e^{4}(j+1)^{2}\Delta^{2}$ and $F_{c}^{2}$ gives $e^{2}\Delta$ up to an overall
constant in the computations (\ref{co5.15}) to (\ref{co5.21}). This clearly
generalizes to arbitrary correlators.

In summary, we have verified that the contractable non-self-intesecting Wilson
loop for $U(1)$ and $SU(2)$ can be computed from our correlators and the
right-hand side of (\ref{co5.3}), at least to $O(\Delta^{4})$. One would think
that a discrepancy would have already appeared at or before $O(\Delta^{4})$,
so that this provides an impressive check of our correlators, and also 
``experimental''
evidence that the Stokes' theorem (\ref{co5.1}) is valid for the 
abelianization gauge, 
even for $SU(2)$\footnote{For $SU(N)$ with $N>2$ the abelian Stokes theorem works up to order $\Delta^{2}$ but seems to fail at next order \cite{steve}.}. A formal proof of this remarkable result is lacking.
Finally, we mention that we have not assumed that the loop $\gamma$ was
infinitesimal, or that $\Delta$ was small. This provides one more advantage of
the abelianization gauge, which seems to provide an enormous simplification as
compared to the non-abelian Stokes' theorem \cite{Stokes}.

\noindent{\bf Acknowledgements}

We wish to thank M. Crescimanno, J. M. Isidro, S. Naculich and H. Riggs for 
helpful discussions.

\appendix{\Large {\bf Appendix A: The Kernel on the Disc}}
\renewcommand{\theequation}{A.\arabic{equation}}
\setcounter{equation}{0}
\newcommand{\R}{\mbox{\rm r}}

In the main text we started with the action for 2d Y.M. coupled to an external
source on the closed surface \s, and then used the abelianization method of
\cite{BT2,BT3} for the path integral, where the nontrivial $U(N)$ and $T$
bundles over \s\ had to be incorporated. Functional differentiation with
respect
to $J$ was then used to derive the correlators. Similar results can be obtained
by a different method. In \cite{BT1} Blau and Thompson computed the kernel
for 2d Y.M. on a disc and used it to find the partition function and Wilson
loops for closed surfaces. One starts with the theory on a disc and sets a
boundary condition for the holonomy of the gauge field around the boundary
$\gamma(t)$ of the disc $D$. The path-ordered exponential representing this
holonomy is expressed in terms of an ordinary exponential, by means of a
functional integration over auxiliary anticommuting fields. This introduces an
explicit $A_{\mu}$ dependence of the integrand in the path integral. However,
for the simple topology of the disc, the Schwinger-Fock gauge is available, and
this allows one to express $A_{\mu}$ in terms of $F_{\mu\nu}$, so that the
Nicolai map becomes useful. The integration over the gauge field and over the
auxiliary fermionic fields can then be performed \cite{BT1}.

When we couple the theory to an external source $J$, the calculation follows
along similar lines. The resulting partition function is,
\begin{equation}
Z_{D}(g,J)=\sum_{\R}\dim(\R)\{\exp(\frac{-e^{2}AC_{2}(\R)}{2})\exp\mbox{$[$}
\frac{e^{2}}{2}\int_{D}d\mu\;\T(J^{2})\mbox{$]$}\;\Char_{\R}(g\cdot P_{t}
\exp(ie^{2}
\oint_{\gamma^{-1}}\rho^{a}_{\mu}
\R^{a}))\}
\label{ap1.1}
\end{equation}
where \R\ denotes an irreducible representation of the gauge group $G$,
$\Char_{\R}$ and $\R^{a}$
are respectively the corresponding Weyl character and generators, $g$ is the
holonomy around $\gamma(t)$, $P_{t}$ is the path-ordering operator and
$\rho_{\mu}^{a}$ is given by
\begin{equation}
\rho_{\mu}^{a}(t)=\int_{0}^{1}sds\epsilon_{\mu\nu}\gamma^{\mu}(t)
\sqrt{g(s\gamma(t))}J^{a}(s\gamma(t))
\label{ap1.2}
\end{equation}
where we have cartesian coordinates $x^{1}$,$x^{2}$ on the disc $D$ with,
\begin{equation}
x^{\mu}(s,t)=s\gamma^{\mu}(t) \makebox[1in]{for} 0\leq s,t\leq 1
\label{ap1.3}
\end{equation}

When glueing together two such discs, by identifying holonomies along
boundaries and
integrating over all such possible holonomies, one should be careful in
specifying the $t=0$ and $t=1$ points of the boundaries consistently.
Results for closed surfaces are obtained by decomposing the holonomy into the
product of the holonomies along the cycles of \s\, and making the corresponding
identifications. One can now perform functional derivatives of (\ref{ap1.1})
with respect to $J$ to compute the field strength correlation functions. The
correlators should be independent of the way one chooses to decompose the
surface \s. So, for example, one can construct a sphere out of two discs, with
boundaries $\gamma(t)$ and $\gamma^{-1}(t)$, or by considering only one disc
with boundary $\alpha\alpha^{-1}(t)$, and the answer should be the same in the
two cases. Let us construct the regularized 2-point function on the sphere, 
considering the
1 disc approach first. One decomposes the holonomy in
to $gg^{-1}$ in
(\ref{ap1.1}) and in this case the integration over holonomies with
$\int_{G}dg$ just produces a factor of unity.
Functional differentiation with respect to $J$ then produces,
\begin{equation}
\langle\xi^{a}(x)\xi^{b}(y)\rangle=\beta\delta^{2}_{x,y}\delta^{ab}+
\frac{1}{Z_{S^{2}}}(-e^{4})\sum_{\R}\dim(\R)\exp(\frac{-e^{2}AC_{2}(\R)}{2})
\Char_{\R}(\R^{a}\R^{b})
\label{ap1.4}
\end{equation}
where $\beta$ is a renormalization dependent constant. Notice this is a
different gauge than
 the one considered in the body of the text.
If we construct the sphere out of two discs the partition function will be,
\begin{equation}
Z_{S^{2}}=\int_{G}dg\;Z_{D}(g,J_{D})\;Z_{D^{'}}(g^{-1},J_{D^{'}})
\label{ap1.5}
\end{equation}
and this integration can be carried out using the familiar orthogonality
relations between Weyl characters,
\begin{equation}
\int_{G}dg\;\Char_{\R}(gh)\Char_{\R^{'}}(g^{-1}h^{'})=\frac{\delta_{\R\R^{'}}}
{\dim(\R)}\Char_{\R}(hh^{'})
\label{ap1.6}
\end{equation}
with the result,
\begin{equation}
Z_{S^{2}}=\sum_{\R}\dim(\R)\exp(\frac{-e^{2}AC_{2}(\R)}{2})\exp(\frac{e^{2}}{2}
\int_{S^{2}}
d\mu\;\T(J^{2}))\;\Char_{\R}(P_{t}\exp(ie^{2}\oint_{\gamma}\rho)P_{t}\exp(
ie^{2}\oint_{\gamma^{'}}\rho^{'}))
\label{ap1.7}
\end{equation}
where $\rho_{\mu}=\sum_{a}\rho^{a}_{\mu}\R^{a}$ and $\rho^{'}$ is obtained from
$J_{D^{'}}$ as in (\ref{ap1.2}). The two-point function obtained from
(\ref{ap1.7})
consists of several terms which depend on wheter or not $x$ and $y$ are in $D$
or
 $D^{'}$. These terms can be understood as arising from glueing the one-point
function on $D$ with the one-point function on $D^{'}$, and the two-point
function
on $D$ with the partition function on $D^{'}$ and vice-versa. The result is the
same as the one in (\ref{ap1.4}), as expected. Notice that in this calculation,
the \TT\ valued fields play no special role, all indices in the Lie algebra
being treated equally. The gauge invariant two-point function, for the
appropriate choice of the renormaliza
tion constant $\beta$, is identical to
the one computed in section 3.

For higher genus surfaces and higher point functions this method becomes
cumbersome. The explicit dependence of the partition function on the choice of
coordinates $s$ and $t$ appears in the correlators, the ordering in which the
generators $\R^{a_{i}}$ appear inside traces being dependent on the ordering of
the coordinates $t(x_{i})$. Thus the symmetry must be restored by hand, and
only
then do the higher order Casimir operators appear. So, for closed surfaces the
method based on abelianization is clearly simpler and more elegant, however
this second method can be applied to the computation of electric field
correlators on surfaces with boundaries.

\appendix{\Large {\bf Appendix B: The Weyl Group Averages}}
\renewcommand{\theequation}{B.\arabic{equation}}
\setcounter{equation}{0}

Here we will present the averages over the
 Weyl group used in section 3.
Recall that the Weyl group acts as the symmetric group of $N$ elements,
$S_{N}$, by permuting the coordinates $(l+\rho)^{i}$. All the identities that
are needed can be deduced from,
\begin{equation}
 \sum_{i=1}^{N}(l+\rho)^{i}=n
\label{ap2.1}
\end{equation}
where $n$ is the $u(1)$ highest weight corresponding to the $U(N)$ irreducible
representation of highest weight $l$. For the two-point
function (\ref{3.2}) we need to compute,
\begin{equation}
 \frac{1}{N!}\sum_
{\sigma}(l+\rho)^{\sigma_{a}}(l+\rho)^{\sigma_{b}}
\label{ap2.2}
\end{equation}
If $a\neq b$ (\ref{ap2.2}) becomes,
$$
 \frac{1}{N(N-1)}\sum_{\sigma_{a}\neq \sigma_{b}}(l+\rho)^{\sigma_{a}}
(l+\rho)^{\sigma_{b}}
$$
and then the sum over $\sigma_{b}$ produces
$$
 \frac{1}{N(N-1)}\sum_{\sigma_{a}}[n(l+\rho)^{\sigma_{a}}-
((l+\rho)^{\sigma_{a}})^{2}]
$$
so that finally we have for the case $a\neq b$
$$
 \frac{1}{|W|}\sum_{\sigma}(l+\rho)^{\sigma_{a}}(l+\rho)^{\sigma_{b}}=
\frac{1}{N(N-1)}[n^{2}
-(l+\rho)^{2}]
$$
(Note we have the normalization  $v^{2}=\sum_{i=0}^{N}v^{i}v^{i}$,
for $v\in \TT$).

Similar reasoning  for $a=b$ gives the result presented in section 3:
\begin{equation}
 \frac{1}{|W|}\sum_{\sigma}(l+\rho)^{\sigma_{a}}(l+\rho)^{\sigma_{b}}
=p^{ab}(l+\rho)^{2}+m^{ab}n^{2}
\label{ap2.5}
\end{equation}
with,
\begin{eqnarray}
  p^{ab}= \left\{ \begin{array}{cc}
                  \frac{-1}{N(N-1)} & \mbox{if $a\neq b$} \\
                  \frac{1}{N}       & \mbox{
if $a=b$}
                  \end{array}
           \right.
\makebox[.9in]{and}
  m^{ab}= \left\{ \begin{array}{cc}
                   \frac{1}{N(N-1)} & \mbox{if $a\neq b$} \\
                   0                & \mbox{if $a=b$}
                  \end{array}
           \right.
\label{ap2.6}
\end{eqnarray}

For the four-point function we have by similar arguments,
\begin{eqnarray}
\frac{1}{|W|}\sum_{\sigma}(l+\rho)^{\sigma_{i}}(l+\rho)^{\sigma_{j}}
(l+\rho)^{\sigma_{k}}(l+\rho)^{\sigma_{l}}= 
\nonumber \makebox[1.5in]{} \\
a^{ijkl}\sum_{p}(l+\rho)_{p}^{4}+b^{ijkl}[(l+\rho)^{2}]^{2}+c^{ijkl}n
\sum_{p}(l+\rho)_{p}^{3}+d^{ijkl}n^{2}(l+\rho)^{2}+e^{ijkl}n^{4}
\makebox[.1in]{}
\label{ap2.7}
\end{eqnarray}
The coefficents of the various Weyl group invariant terms are completely
symmetric in the indices $i$,$j$,$k$,$l$. With
\[ \varepsilon^{ij}= \left\{ \begin{array}{cc}
                             1 & \mbox{$i=j$} \\
                             0 & \mbox{$i\neq j$}
                       
      \end{array}
                      \right. \]
we have (with no sum on repeated indices in (\ref{ap2.8})):
\begin{eqnarray}
a^{ijkl}= \left\{ \begin{array}{cc}
                  \frac{-6(N-4)!}{N!} & \A \\
                  \frac{2(N-3)!}{N!}  & \B \\
                  \frac{-(N-2)!}{N!}  & \C \\
                  \frac{-(N-2)!}{N!}  & \E \\
                  \frac{(N-1)!}{N!}   & \D
                  \end{array}
           \right.
\makebox[.3in]{}
b^{ijkl}= \left\{ \begin{array}{cc}
     
             \frac{3(N-4)!}{N!} & \A \\
                  \frac{-(N-3)!}{N!} & \B \\
                  0                   & \C \\
                  \frac{(N-2)!}{N!}  & \E \\
                  0                   & \D
                  \end{array}
          \right.
\nonumber
\\
c^{ijkl}= \left\{ \begin{array}{cc}
                  \frac{8(N-4)!}{N!}  & \A \\
                  \frac{-2(N-3)!}{N!} & \B \\
                  \frac{(N-2)!}{N!}   & \C \\
                  0                   & \mbox{otherwise}
                  \end{array}
           \right.
\nonumber
\makebox[.3in]{}
d^{ijkl}= \left\{ \begin{array}{cc}
                  \frac{-6(N-4)!}{N!} & \A \\
                  \frac{(N-3)!}{N!}   & \B \\
                  0                    & \mbox{otherwise}
                  \end{array}
            \right.
\\
e^{ijkl}= \left\{ \begin{array}{cc}
                  \frac{(N-4)!}{N!} & \A \\
                  0                 & \mbox{otherwise}
                  \end{array}
            \right.
\makebox[2in]{}
\label{ap2.8}
\end{eqnarray}
Notice that when one contracts with the generators of the maximal torus in
(\ref{3.4}), only the $i=j=k=l$ case contributes.


\end{document}